\documentclass[epjCONF,columns]{svjour} 
\usepackage{graphics}
\usepackage[varg]{txfonts} 
\usepackage[latin1]{inputenc}
\usepackage[numbers, sort&compress]{natbib}

\session-title{FUSION14}
\begin{document}
\title{Microscopic study of the effect of intrinsic degrees of freedom on fusion}
\author{C. Simenel\inst{1}\fnmsep\thanks{\email{cedric.simenel@anu.edu.au}} \and M. Dasgupta\inst{1} \and D. J. Hinde\inst{1} \and V. E. Oberacker\inst{2} \and A. S. Umar\inst{2} \and E. Williams\inst{1}}
\institute{Department of Nuclear Physics, RSPE, Australian National University, Canberra, Australian Capital Territory 0200, Australia \and Department of Physics and Astronomy, Vanderbilt University, Nashville, Tennessee 37235, USA}
\abstract{
Fusion cross-sections are computed for the $^{40}$Ca$+^{40}$Ca system over a wide energy range with two microscopic approaches where the only phenomenological input is the Skyrme energy density functional. 
The first method is based on the coupled-channels formalism, using the bare nucleus-nucleus potential calculated with the frozen Hartree-Fock technique and the deformation parameters of vibrational states computed with the  time-dependent Hartree-Fock (TDHF) approach.
The second method is based on the density-constrained TDHF method to generate nucleus-nucleus potentials from TDHF evolution. 
Both approaches incorporate the effect of couplings to internal degrees of freedoms in different ways.
The predictions are in relatively good agreement with experimental data.} 
\maketitle
\section{Introduction}
\label{intro}

Near-barrier fusion can be strongly affected by the coupling between relative motion and internal degrees of freedom of the collision partners \cite{das85,das98}.
In particular,  couplings to rotational states \cite{lei95}, as well as to low-lying collective vibrations \cite{mor94,ste95a,das98,bal98}, can  enhance sub-barrier fusion by orders of magnitude as compared to a single barrier penetration model. 
Indeed, the couplings to collective states induce a dynamical change of the density and thus different potential barriers can be present in the entrance channel. 
In addition to generate a barrier distribution, the couplings generally also shift the centroid of this distribution, making it difficult to determine the bare (i.e., without effects of the couplings) nucleus-nucleus potential. 

Time-dependent Hartree-Fock (TDHF) calculations have shown that the effects of the couplings on fusion are expected to disappear at high energy as the shapes of the nuclei do not have time to change during the approach \cite{was08}. 
In this case, capture occurs in the bare potential where the nuclei still have their ground-state densities.
As a result, the couplings are expected to induce an energy dependence of the potential \cite{was08,sim08,obe10,uma14}.

The coupled-channels (CC) method is the standard approach to investigate the effect of couplings on fusion  \cite{lin84,das85,tho85,esb87,ste90,hag12}.
CC calculations require external parameters to describe the nucleus-nucleus potential and the couplings to internal degrees of freedom, such as energies and deformation parameters of collective states. 
The latter have been often measured for stable nuclei (see, e.g., Refs~\cite{jun11} and \cite{kib02} for compilations of $2^+_1$ and $3^-_1$ states, respectively), and nucleus-nucleus potential parametrisations such as the Aky\"{u}z and Winter \cite{aky79} or the Sao-Paulo \cite{cha02} potentials have been shown to reproduce reasonably well near-barrier fusion cross-sections. 

The application of the CC method to reactions with exotic radioactive beams will be more problematic.
Indeed, little is usually known about the structure of exotic nuclei. 
In addition, it is not clear whether or not standard parametrisations of nucleus-nucleus potentials could be applied to exotic nuclei, in particular close to the drip lines, where neutron or proton skins and halos could be present. 

In this contribution, we discuss a recently proposed method \cite{sim13b} where the nucleus-nucleus potential and the properties of the collision partners entering CC calculations are determined from purely microscopic calculations with the Hartree-Fock (HF) method and its time-dependent extension (TDHF). 
In this method, the only inputs are the choice of the states to be coupled and the Skyrme energy density functional (EDF) \cite{sky56} describing the phenomenological  interaction between the nucleons. 
It is worth noting that the parameters of the Skyrme EDF are fitted to structure properties only (see, e.g., \cite{cha98}).
The resulting fusion cross-sections calculations are then computed without any input coming from reaction mechanisms. 
In this work, the $^{40}$Ca$+^{40}$Ca system is considered as a simple benchmark of this method. 

Another method is also used to investigate the effect of couplings on fusion in this system. 
It is based on the  density-constrained TDHF (DC-TDHF) technique to extract the nucleus-nucleus potential from TDHF trajectories \cite{uma06}.
In this approach, the couplings induce an energy dependence of the potential. 

A brief outline of the first method is presented in section~\ref{sec:method}.
TDHF and CC calculations are described in section~\ref{sec:CC} and compared with experimental data.
The energy dependence of the nucleus-nucleus potential is then investigated in section~\ref{sec:Edep} with DC-TDHF before to conclude in section~\ref{sec:conclusion}.

\section{Method \label{sec:method}}

We focus on the effect on the fusion process of the coupling to { vibrational} states.
The only input of the present method is the Skyrme effective interaction \cite{sky56}. 
The basic steps of the approach are:
\begin{enumerate} 
\item The bare nucleus-nucleus potential is computed from the frozen Hartree-Fock technique. 
\item A TDHF code is used to compute the strength function of vibrational modes using the linear response theory.
\item The strength function is used to extract the energy and deformation parameter of collective vibrational states.
\item The bare nucleus-nucleus potential and the parameters of the coupling are used in standard coupled-channels calculations to determine fusion cross-sections. 
\end{enumerate}

Near-barrier TDHF calculations are also used to determine the fusion threshold which provides a realistic estimate of the centroid of the barrier distribution.
If the centroid of the final barrier distribution obtained from CC calculations is in good agreement with the TDHF fusion threshold, then we can reasonably conclude that the most relevant internal degrees of freedom have been included in the CC calculations. 
More details on the method can be found in Ref.~\cite{sim13b}.

\section{TDHF and CC calculations \label{sec:CC}}

\begin{figure}[ht]
\resizebox{1.0\columnwidth}{!}{
\includegraphics{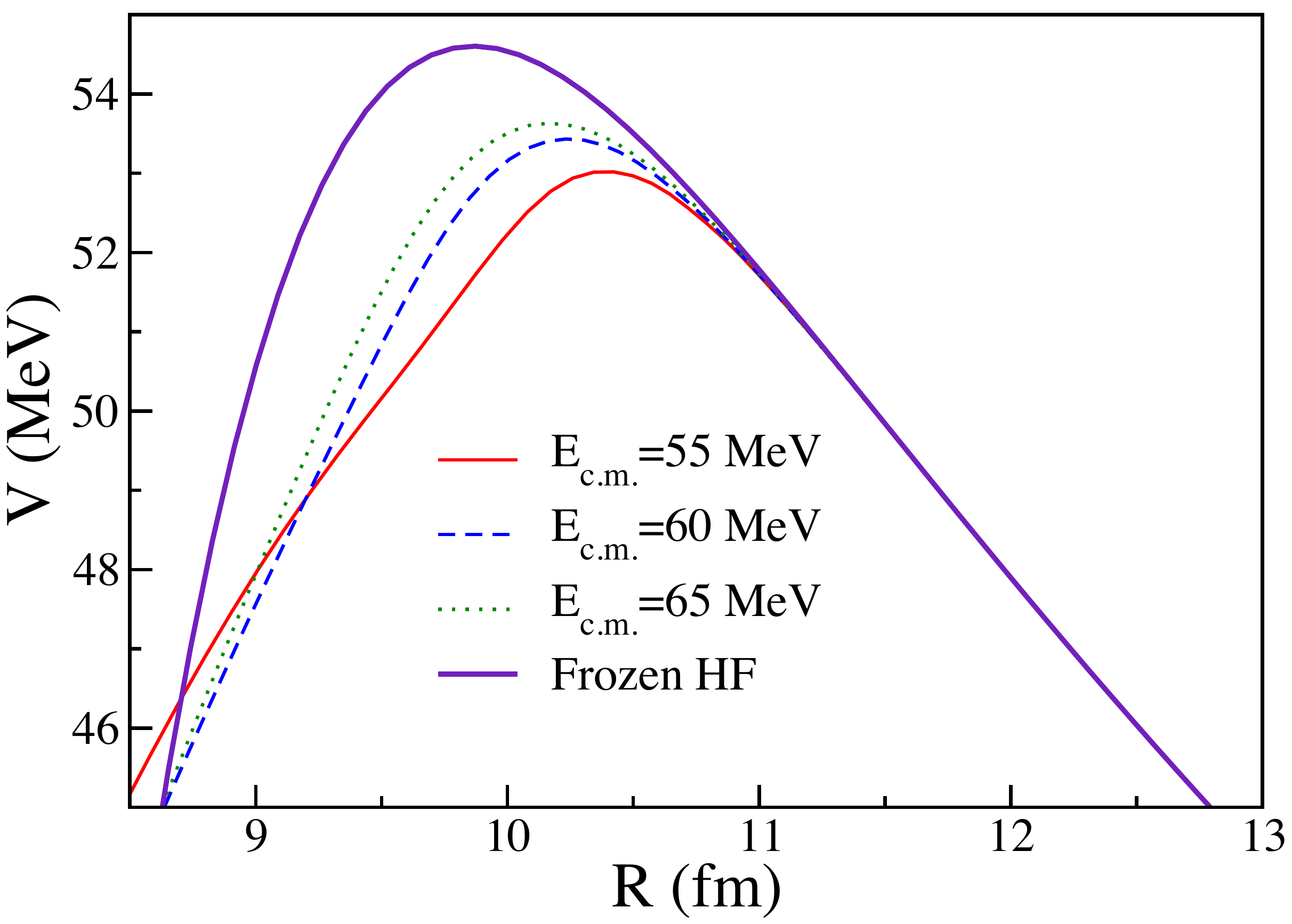}}
\caption{\label{fig:Pot}  
Total potential energy as a function of the relative distance between the fragments in $^{40}$Ca$+^{40}$Ca central collisions. The frozen HF potential is represented by a thick solid line. DC-TDHF potentials calculated from TDHF density evolutions at bombarding energies $E_{c.m.}=55$~MeV (thin solid line), 60~MeV (dashed line), and 65~MeV (dotted line) are also shown.}
\end{figure}

The potential to be used in CC calculations is computed with the frozen HF method \cite{sim08,was08,sim12} with the SLy4$d$ Skyrme functional \cite{kim97}.
It is plotted for the $^{40}$Ca$+^{40}$Ca system in Fig.~\ref{fig:Pot} with a thick line.
The resulting barrier height is $V_B=54.6$~MeV at $R_B=9.9$~fm. 

\begin{figure}[ht]
\resizebox{1.0\columnwidth}{!}{
\includegraphics{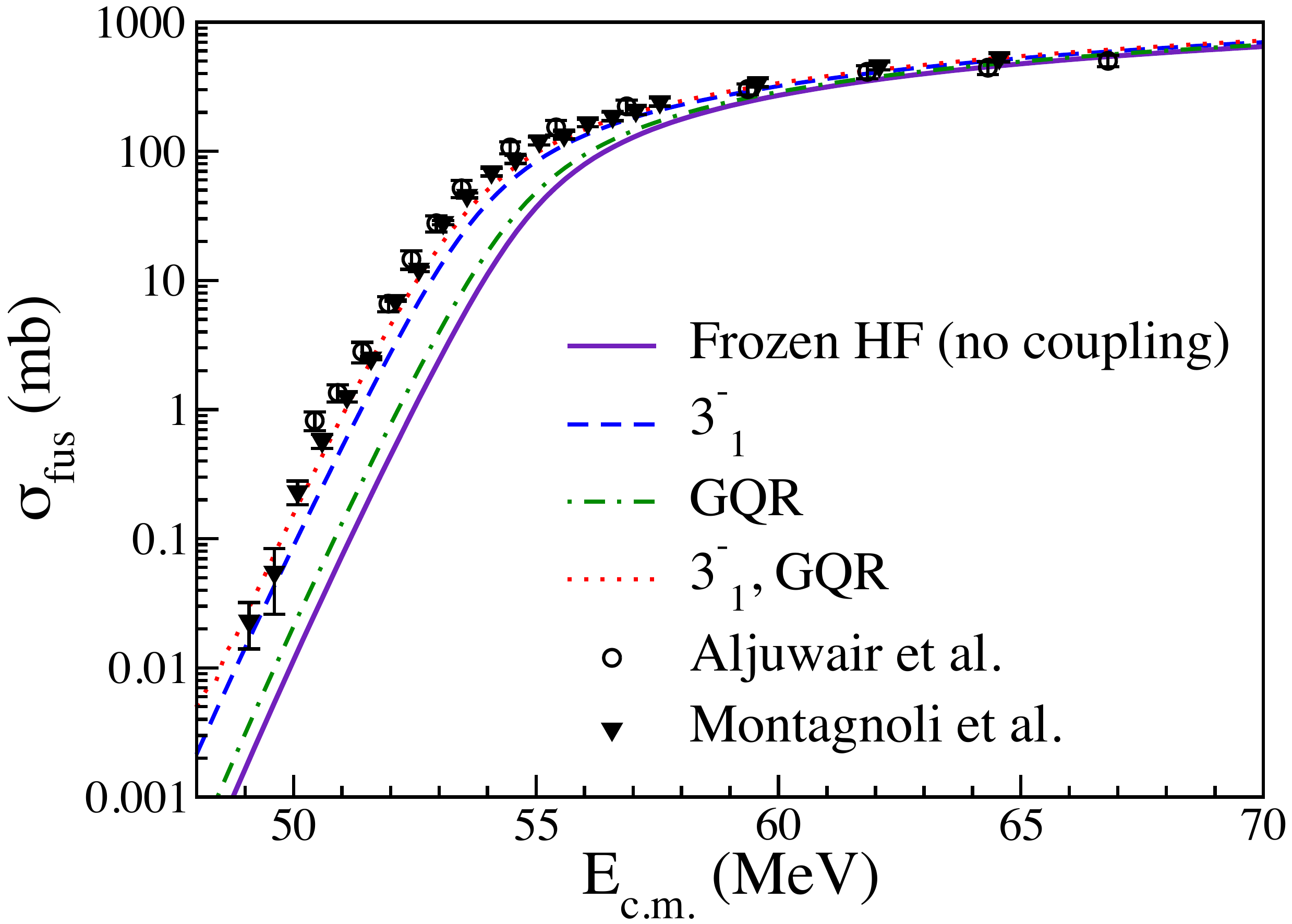}}
\caption{\label{fig:sigma_CaCa_log}  
Fusion cross-sections on logarithmic scale as a function of center of mass energy for the $^{40}$Ca$+^{40}$Ca system using the frozen HF potential. The thick solid line shows the results without coupling. Couplings to the $3^-_1$ state and to the GQR lead to the cross sections plotted with the dashed line and with the dotted-dashed line, respectively, and to the dotted line when both states are included in the coupled-channels calculations. The data from Aljuwair {\it et al.} \cite{alj84} and the more recent ones from Montagnoli {\it et al.} \cite{mon12} are plotted with open circles and filled triangles, respectively.}
\end{figure}

\begin{figure}[ht]
\resizebox{1.0\columnwidth}{!}{
\includegraphics{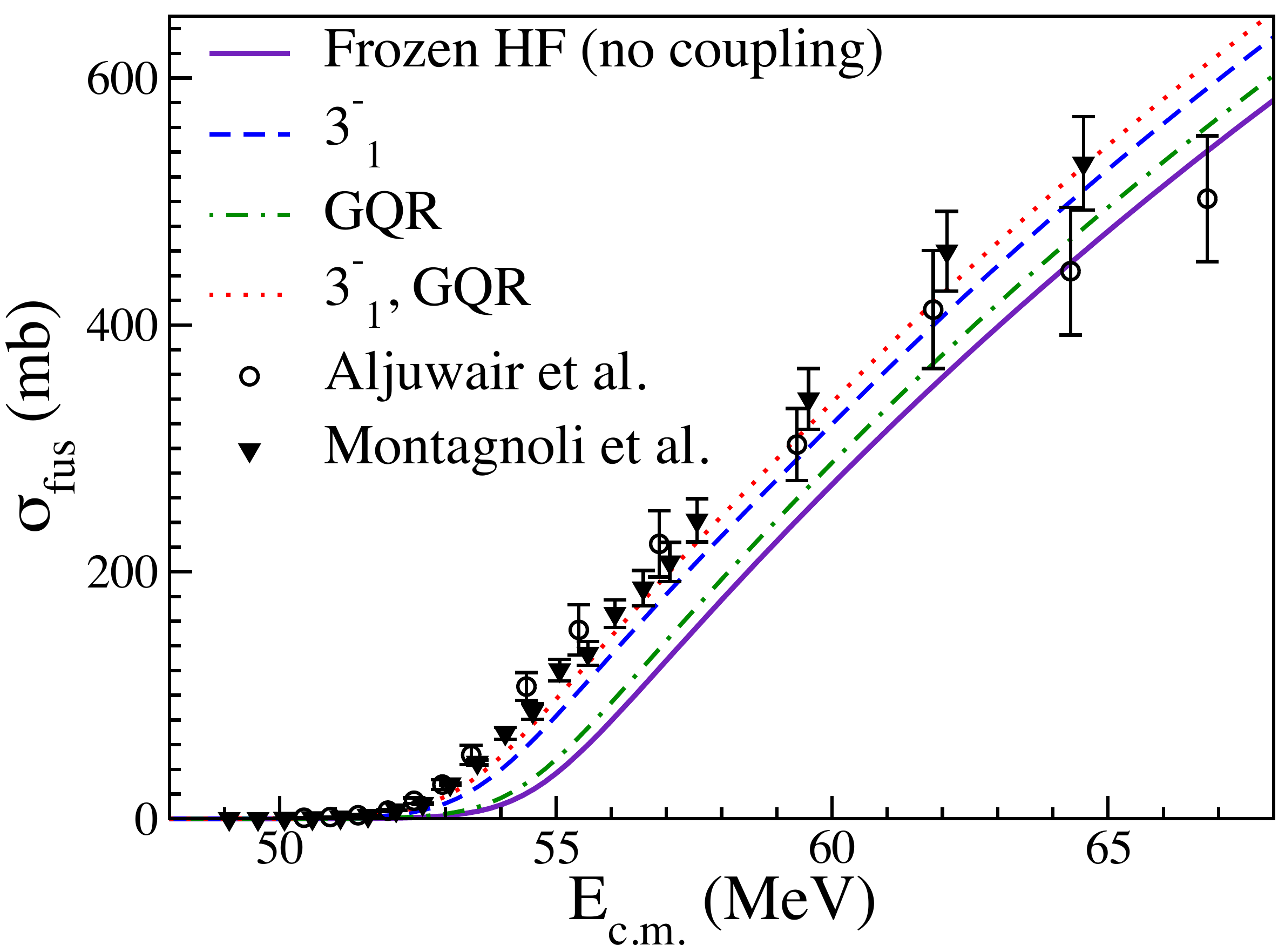}}
\caption{\label{fig:sigma_CaCa_lin}  
Same as Fig.~\ref{fig:sigma_CaCa_log} in linear scale.}
\end{figure}

\begin{figure}[ht]
\resizebox{1.0\columnwidth}{!}{
\includegraphics{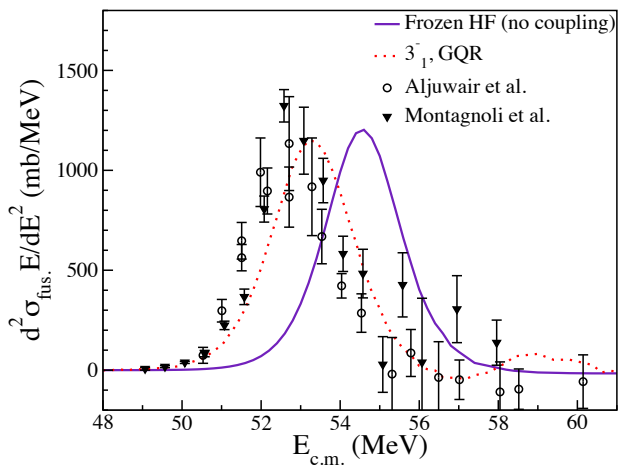}}
\caption{\label{fig:BD}  
Experimental fusion barrier distribution as a function of center of mass energy for the $^{40}$Ca$+^{40}$Ca system.
The lines show the results obtained with the frozen HF potential without couplings (solid line) and with couplings to the $3^-_1$ and GQR states (dotted line). Experimental data are  shown with symbols \cite{alj84,mon12}.}
\end{figure}

A first guess of the fusion cross-sections $\sigma_{fus}$ can be obtained with this potential using the one-barrier penetration model. 
The results are shown with solid lines in Figs.~\ref{fig:sigma_CaCa_log} and~\ref{fig:sigma_CaCa_lin} on logarithmic and linear scales, respectively. 
We see that the calculations strongly underestimate the experimental data.
The experimental barrier distribution, obtained from the second derivative of $\sigma_{fus}E$ \cite{row91}, is plotted in Fig.~\ref{fig:BD}. 
We see that the underestimation of the fusion cross-sections is due to an apparent overestimation of the barrier. 
Of course, this is because couplings are not yet included. 
Indeed, it is well known that  couplings may induce a renormalisation of the potential \cite{hag97}.

To estimate the importance of the vibrational couplings on the fusion cross-sections, we need to determine the properties (energy and deformation parameters) of the vibrational states. 
If available, the latter can be obtained from experimental data, or, alternatively, from theoretical calculations. 
For consistency, we extract these quantities from strength functions computed with a TDHF code (see, e.g., Refs.~\cite{sim12,sim13b} for details of the calculations) using the SLy4$d$ Skyrme functional \cite{kim97}. 
Note that this approach is fully equivalent to the random phase approximation (RPA) which is a standard tool to investigate nuclear vibrations. 

\begin{figure}[ht]
\resizebox{1.0\columnwidth}{!}{
\includegraphics{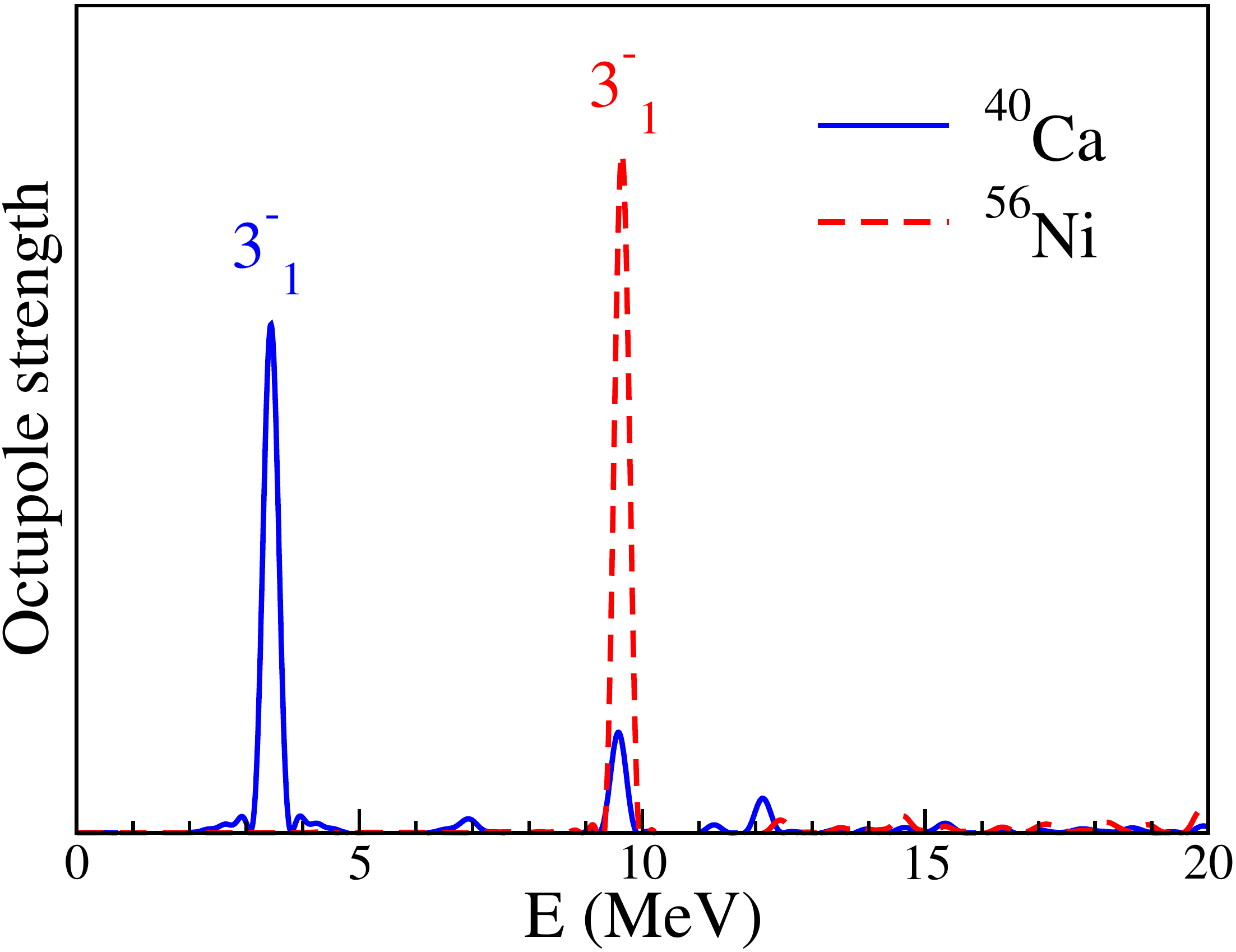}}
\caption{\label{fig:Q3}  
Octupole strength distribution calculated from TDHF response to an octupole excitation in the linear regime.
Results are shown for the $^{40}$Ca (solid line) and $^{56}$Ni (dashed line) doubly magic nuclei.}
\end{figure}

The coupling to octupole vibrations is known to have an important effect on fusion \cite{row10}.
Such coupling is naturally present in time-dependent self-consistent  approaches such as TDHF \cite{sim13b,sim13a}.
Figure~\ref{fig:Q3} shows the  strength function for octupole vibrations in $^{40}$Ca (solid line). 
The main peak at low energy corresponds to the collective $3^-_1$ state. 
It is found at 3.44~MeV, which is reasonably close to the experimental value of 3.74~MeV \cite{kib02}.
Other peaks are also observed at higher energies. 
However the latter are weaker and the main effect on the cross sections is expected to come from the coupling to the low-lying $3^-_1$ state. 
It is interesting to note that this microscopic approach reproduces features such as the fact that the magic number 28 induces a larger energy of the $3^-_1$ state (see dashed line in Fig.~\ref{fig:Q3} showing the octupole strength function  of $^{56}$Ni). 

Other modes of vibrations can also be studied with this technique. 
In particular, the coupling to low-lying  $2^+$ states associated to collective quadrupole vibrations are known to affect near-barrier fusion \cite{mor94,ste95}.
However, the low-lying $2^+$ states of $^{40}$Ca are not found to be collective and the quadrupole strength is essentially located in the giant quadrupole resonance (GQR). 
The TDHF calculations predict the energy of GQR to be $\sim18.1$~MeV.

The deformation parameters $\beta_\lambda$ of vibrational states need also to be determined.
They are directly proportional to the area of their associated peak in the strength function and can then be directly extracted from TDHF calculations \cite{sim13b,sca13}.
We get $\beta_3=0.24$  for the $3^-_1$ state  of $^{40}$Ca.
Note that the experimental $\beta_3=0.3-0.4$ of the $3^-_1$ is larger \cite{kib02} and could then affect more strongly near barrier fusion. 
A coupling strength of $\beta_2=0.16$ is also obtained for the GQR. 
Note that, although the direct decay of giant resonances could be studied with TDHF \cite{cho87,pac88,ave13}, we treat the GQR as a bound state for simplicity.

The effect of the coupling to the $3^-_1$ state and to the GQR in $^{40}$Ca+$^{40}$Ca fusion  is investigated with coupled-channels calculations using the \textsc{ccfull} code \cite{hag99}.
The resulting cross-sections are plotted in Figs.~\ref{fig:sigma_CaCa_log} and \ref{fig:sigma_CaCa_lin}.
We see that the coupling to the $3^-_1$ state accounts for most of the enhancement of the cross-sections as compared to the calculations without couplings. 
However, the effect of the GQR is not negligible and a good agreement with data is obtained when both states are included. 

\begin{figure}[ht]
\resizebox{1.0\columnwidth}{!}{
\includegraphics{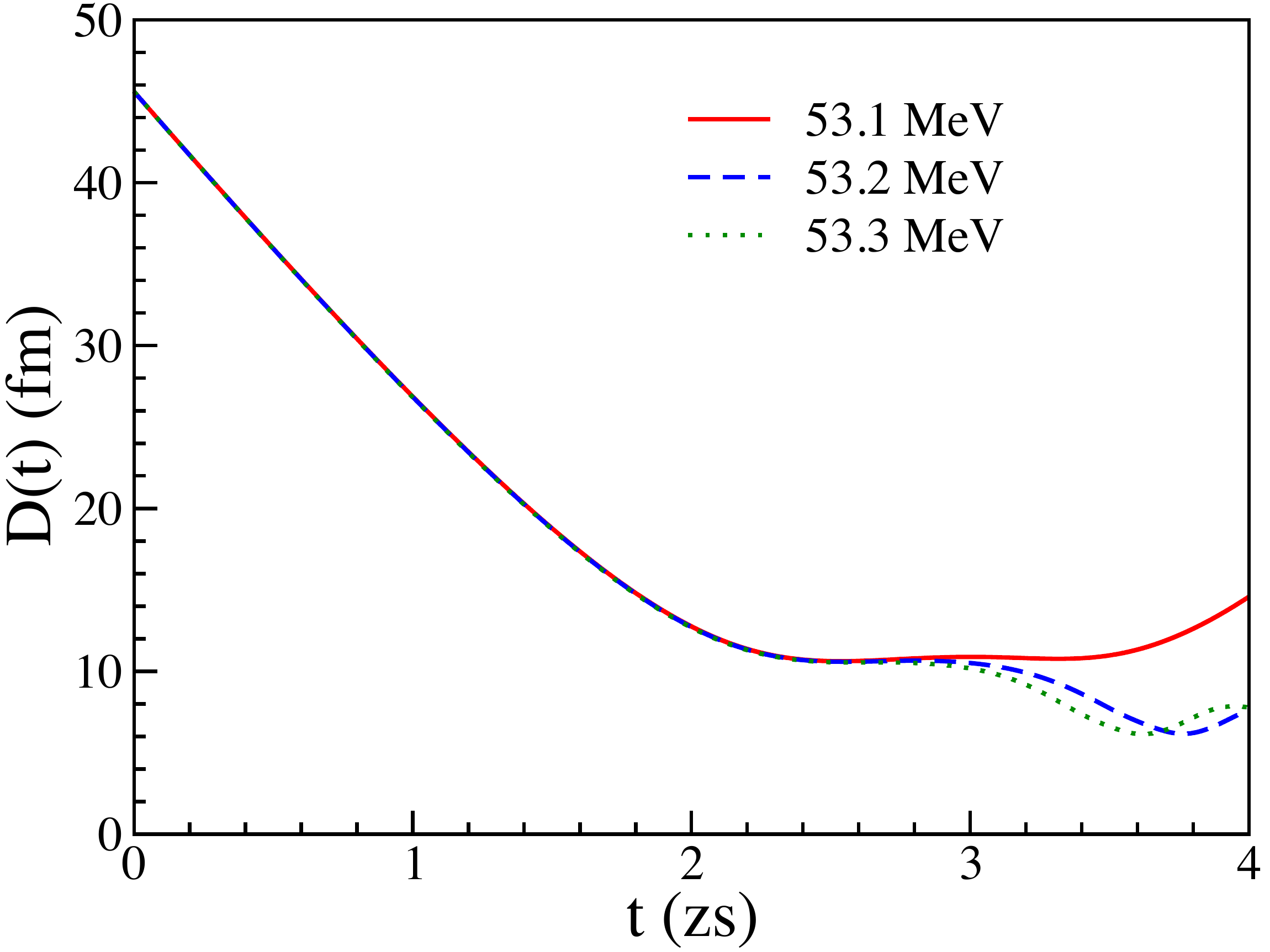}}
\caption{\label{fig:traj}  
Time evolution of the distance between the fragments in $^{40}$Ca$+^{40}$Ca central collisions at $E_{c.m.}=53.1$~MeV (solid line), 53.2~MeV (dashed line) and 53.3~MeV (dotted line).}
\end{figure}

In such a light system, the effect of these couplings is essentially to renormalise the potential to lower energies, as shown by the barrier distribution represented by the dotted line in Fig.~\ref{fig:BD}. 
We see that the experimental barrier distribution is at slightly lower energy, indicating that couplings to other states might play a role. 
This is confirmed by computing the TDHF fusion threshold from the fragment trajectories shown in Fig.~\ref{fig:traj} which lead to a fusion threshold of $53.15\pm0.05$~MeV. 
This threshold is $\sim0.15$~MeV lower than the centroid of the barrier distribution from coupled-channels calculations with couplings to the $3^-_1$ state and to the GQR.
The excitation of other states, such as the giant dipole (GDR) and monopole (GMR) resonances, could be responsible for this small difference. 

\begin{figure}[ht]
\resizebox{0.9\columnwidth}{!}{
\includegraphics{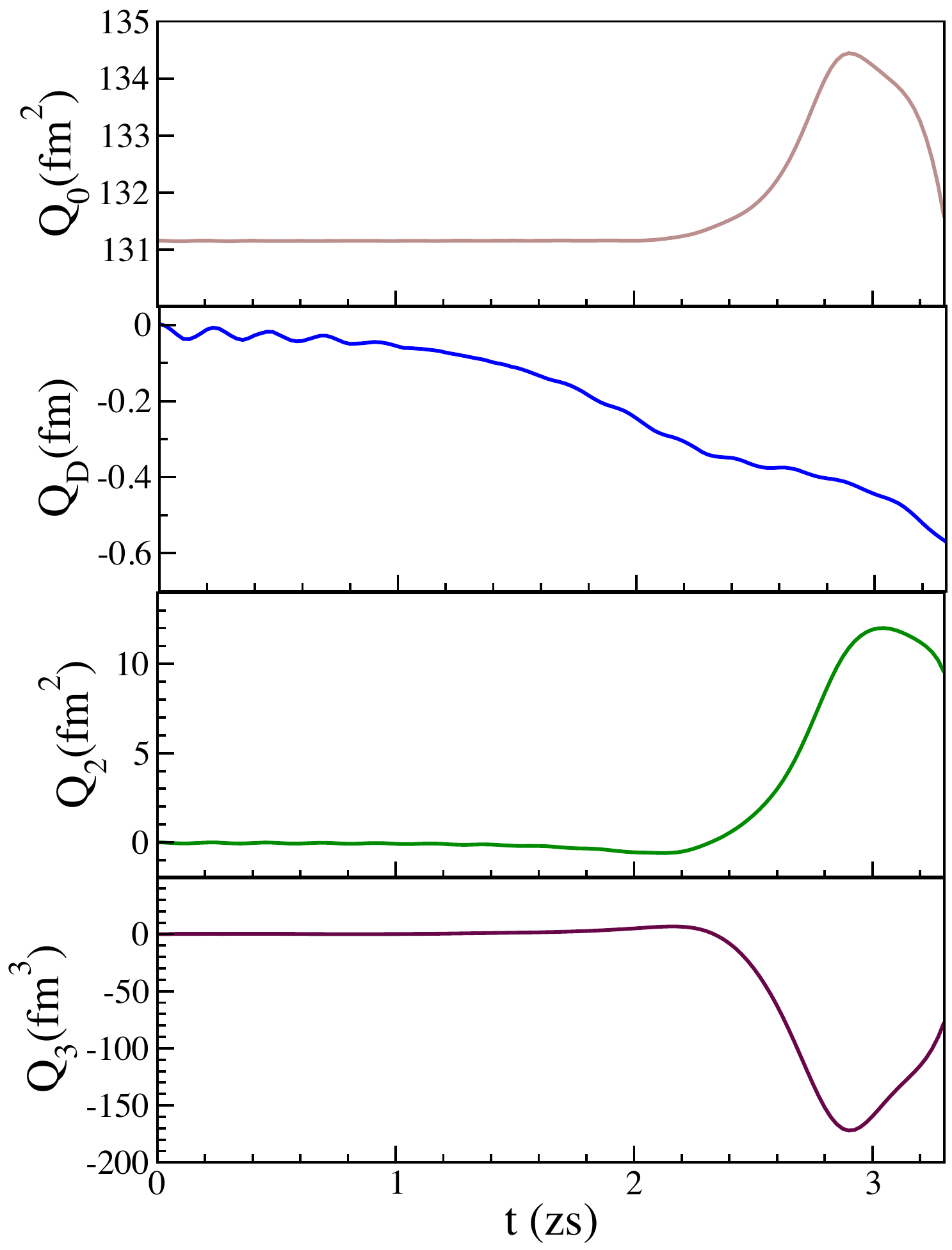}}
\caption{\label{fig:moments}  
Evolution of the monopole ($Q_0$), isovector dipole ($Q_D$), quadrupole ($Q_2$) and octupole ($Q_3$) moments  along the collision axis $x$ and in the region $x>0$ as a function of time for a $^{40}$Ca$+^{40}$Ca central collision at $E_{c.m.}=53.3$~MeV.}
\end{figure}

In order to get a deeper insight into the possible role of the couplings to other modes, the time evolution of different multipole moments of the fragments have been computed.
The monopole, (isovector) dipole, quadrupole and octupole moments of the fragments  in the approach phase are shown in Fig.~\ref{fig:moments} from top to bottom, respectively.
We see that they all deviate from their initial value, indicating polarisation effects which could be interpreted as an effect of couplings.
It is interesting to note that the isoscalar moments  remain essentially unchanged until later times when the nuclear interaction between the fragments become non-negligible. 
This is not the case with the isovector dipole moment which is  affected by long-range Coulomb polarisation. 
The effects of the excitation of the GDR and of the GMR on fusion cross-sections remain to be studied with coupled-channels calculations. 

\section{Energy-dependence of the potential \label{sec:Edep}}

One effect of the couplings is to induce a dynamical change of the density of the collision partners, and, then, of their associated nucleus-nucleus potential. 
It is clear that this effect is intrinsically time dependent. 
For instance, it was shown in Ref.~\cite{sim13b} that the couplings to the $3^-_1$ state in $^{40}$Ca$+^{40}$Ca near the barrier induces an octupole shape of the reactants within a time scale of approximatively one zeptosecond. 
At energies well above the barrier, however, the reaction is more rapid and the density of the collision partners does not have time to deviate from the ground state density.
In particular, this was shown for several systems, including $^{40}$Ca$+^{40}$Ca, by Washiyama and Lacroix with TDHF calculations \cite{was08}. 

This effect can be investigated by extracting nucleus-nucleus potentials directly from TDHF trajectories at different energies. 
Different approaches have been developed in the past to calculate these potentials, such as the dissipative-dynamic TDHF \cite{was08} and the density-constrained TDHF \cite{uma06} methods.
An energy dependence of the potential is usually observed \cite{was08,obe10}.
At near barrier energies, a dynamic adiabatic potential with a barrier modified by the couplings is observed, while at high energy (typically twice the barrier energy \cite{was08}), the bare nucleus-nucleus potential is recovered.
The latter can be estimated with the frozen HF method.

The energy-dependence of the potential is illustrated in Fig.~\ref{fig:Pot} \cite{uma14}.
The potentials calculated with the DC-TDHF method at three TDHF energies are shown.
The TDHF evolutions are computed with the three-dimensional code of Ref.~\cite{uma06a} using the SLy4 interaction \cite{cha98}.  
We observe that the barrier height decreases, and the barrier radius increases, with decreasing bombarding energy. 
The shape of the barrier is also modified at the lowest energies which can have an important effect on deep sub-barrier fusion \cite{uma14}. 

\begin{figure}[ht]
\resizebox{1.0\columnwidth}{!}{
\includegraphics{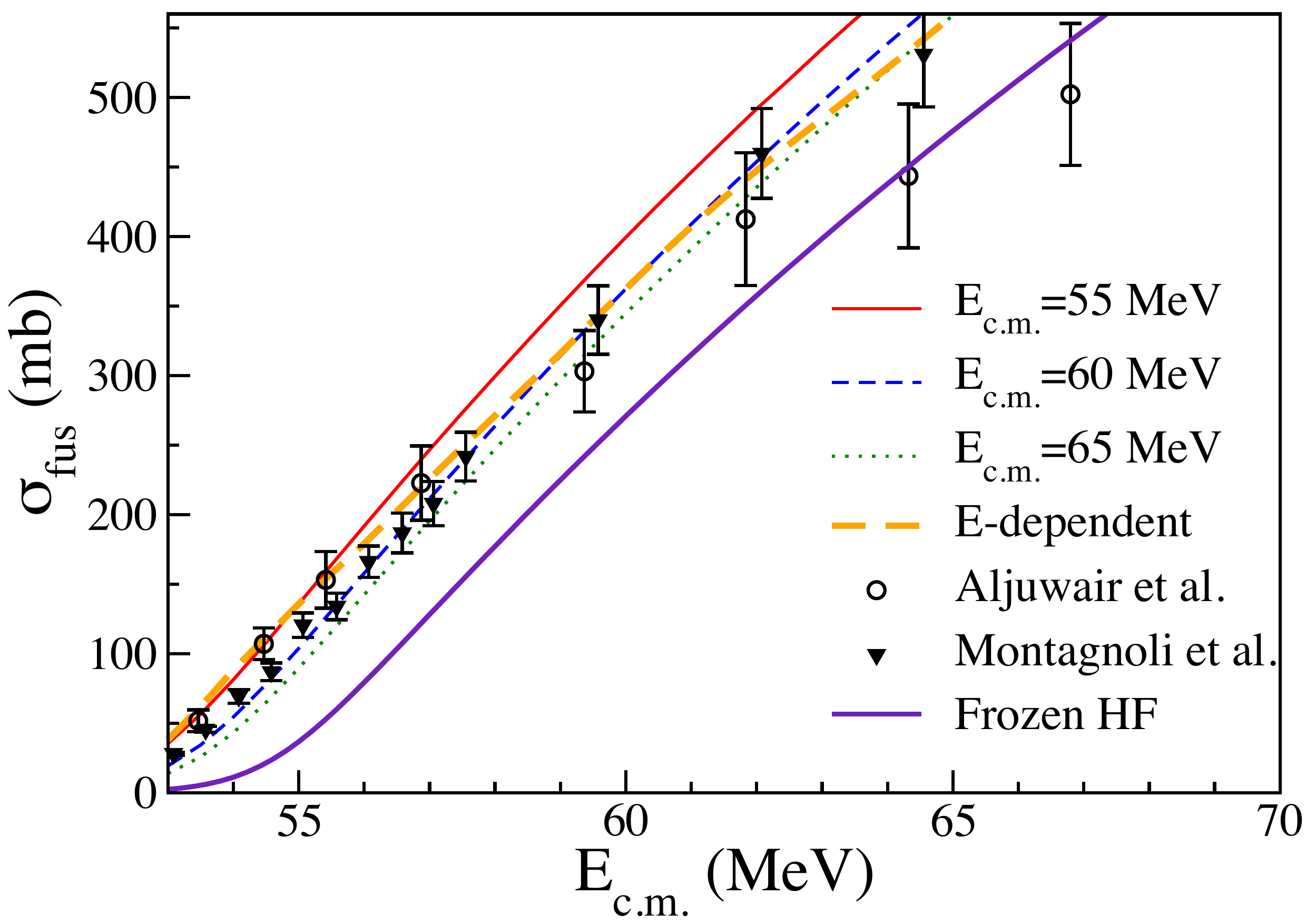}}
\caption{\label{fig:cs_lin}  
Fusion cross-sections (liner scale) for $^{40}$Ca$+^{40}$Ca. Cross-sections obtained from the frozen HF potential and neglecting couplings are shown with the thick solid line. Cross-sections obtained from DC-TDHF potentials calculated with TDHF density evolutions at bombarding energies $E_{c.m.}=55$~MeV, 60~MeV, and 65~MeV are plotted with thin solid, dashed and dotted lines, respectively. The thick dashed line, labelled ''$E-$dependent'', is obtained by combining DC-TDHF calculations at different bombarding energies (see text). Experimental data are shown with symbols \cite{alj84,mon12}.}
\end{figure}

\begin{figure}[ht]
\resizebox{1.0\columnwidth}{!}{
\includegraphics{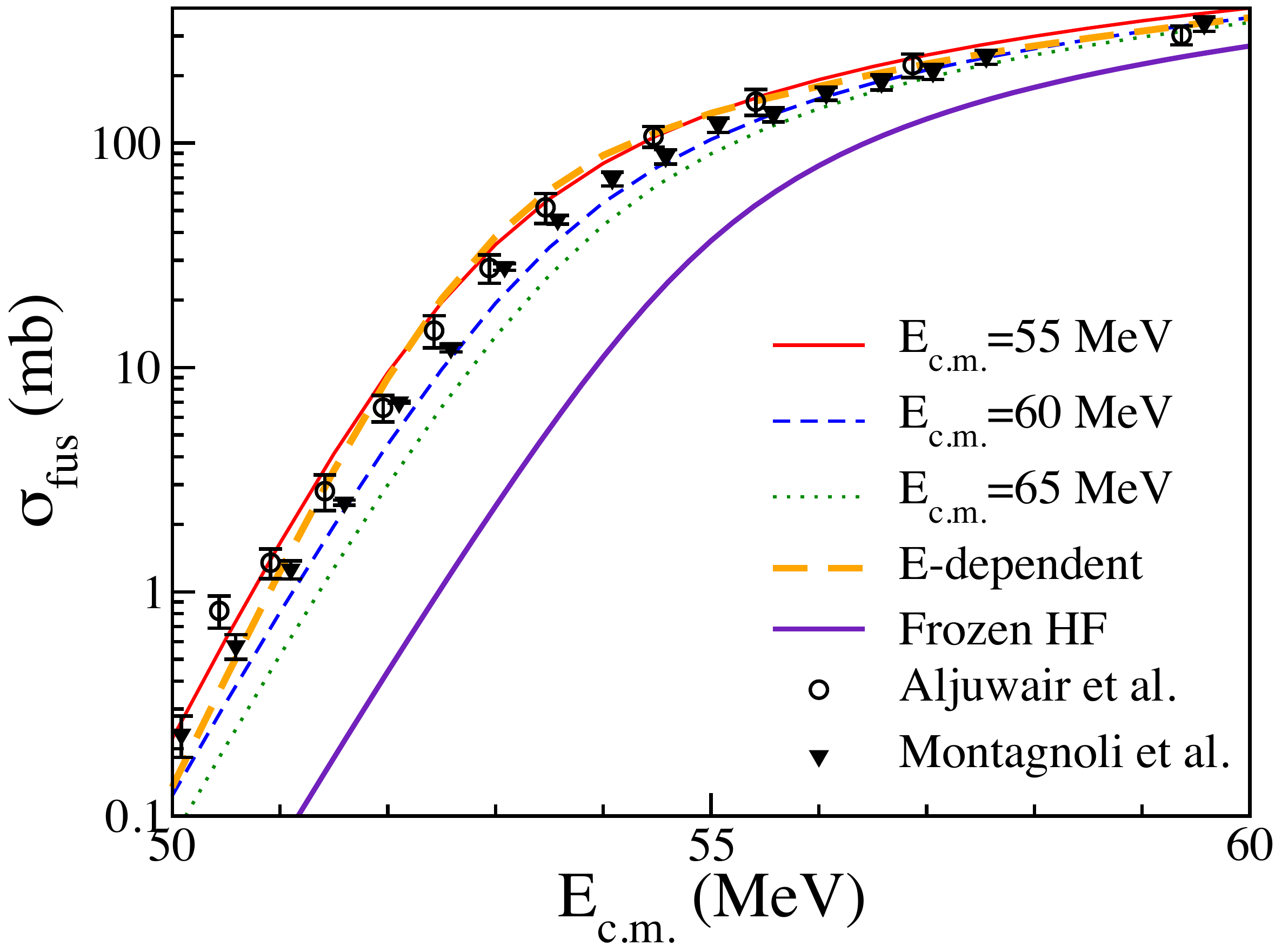}}
\caption{\label{fig:cs_log}  
Same as Fig.~\ref{fig:cs_lin} on logarithmic scale.}
\end{figure}

Each of these potentials can be used in a simple one-barrier penetration model to compute fusion cross-sections. 
These cross-sections are shown in Figs.~\ref{fig:cs_lin} (linear scale) and~\ref{fig:cs_log} (logarithmic scale). 
Of course, these cross-sections are expected to be valid at bombarding energies close to the TDHF energy used to compute the nucleus-nucleus potential. 
Indeed, comparing with experimental data, we clearly see that the potential obtained at bombarding energy $E_{c.m.}=55$~MeV overestimates the data well above the barrier, while the one at $E_{c.m.}=65$~MeV underestimates the cross-sections below the barrier. 

A combined set of cross-sections (thick dashed lines in Figs.~\ref{fig:cs_lin} and \ref{fig:cs_log}) has been determined using potentials extracted from TDHF calculations between $E_{c.m.}=53$~MeV and 65~MeV, in energy step of 1~MeV. Cross-sections below $53$~MeV are computed from the potential obtained at bombarding energy $E_{c.m.}=53$~MeV. 
Indeed, at lower energy TDHF calculations do not lead to fusion and the potential cannot be calculated.  
Despite fluctuations in the experimental data, we see that the calculated cross-sections are in relatively good agreement with data over a large energy range, from well below to well above the barrier. 

\begin{figure}[ht]
\resizebox{1.0\columnwidth}{!}{
\includegraphics{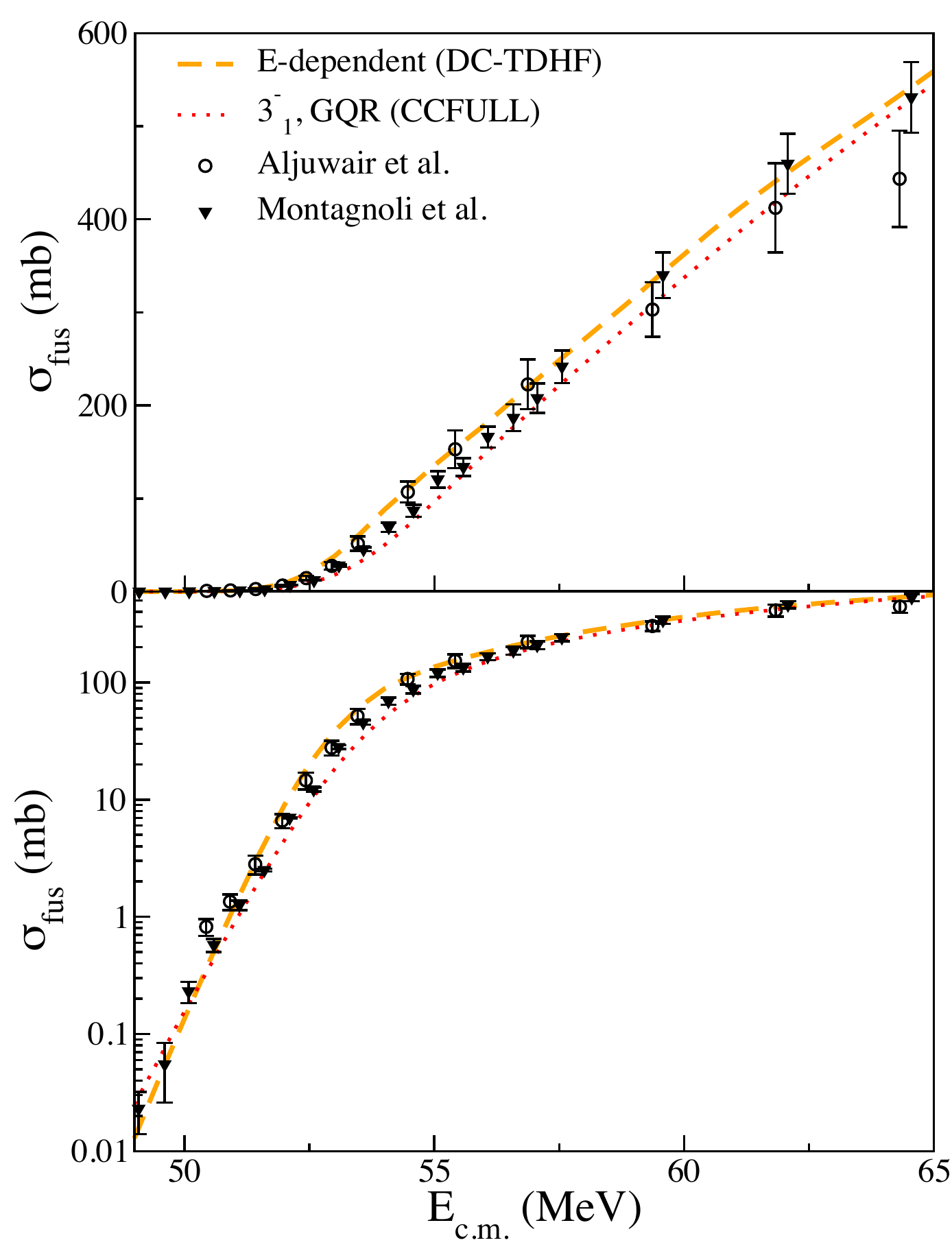}}
\caption{\label{fig:compare}  
Fusion cross-sections on linear (top) and logarithmic (bottom) scales for $^{40}$Ca$+^{40}$Ca as a function of center of mass energy. 
DC-TDHF results (dashed line) obtained from  TDHF calculations at different energies are compared with the coupled-channels results with couplings to the $3^-_1$ and GQR states. Experimental data are shown with symbols \cite{alj84,mon12}.}
\end{figure}

It is interesting to compare the combined DC-TDHF fusion cross-sections with those calculated with the coupled-channels approach. 
Although both methods are very different, they both incorporate, to some extent, the effect of couplings on fusion. 
This comparison is made in Fig.~\ref{fig:compare}.
Overall, they both lead to a reasonable agreement with data over a wide energy range. 
However, we note that, around the barrier, the cross-sections are overestimated by the DC-TDHF calculations and underestimated by the CC ones. 
The underestimation of the cross-sections by the CC calculations could be a signature that more couplings should be included. 
In principle, the DC-TDHF calculations include all couplings, but only in an approximated way.
Indeed, in this approach, only one  barrier, including the effect of the couplings ''on average'', is present at each energy.
It is then not surprising that, at near barrier energies where the couplings could induce structure in the barrier distribution, the DC-TDHF cross-sections are not in perfect agreement with data. 

\section{Conclusions \label{sec:conclusion}}

Two methods have been used to predict fusion cross-sections in the $^{40}$Ca$+^{40}$Ca system over a wide energy range from well below to well above the barrier. 
In both cases, the only inputs are the parameters of the Skyrme energy density functional. 
Both methods lead to a relatively good agreement with experimental data.

The first method is based on the coupled-channels formalism where both the nucleus-nucleus potential and the coupling parameters are computed using the TDHF approach.
It confirms the importance of the low-lying octupole states. 
The GQR also induces a small renormalisation of the potential. 
The role of other giant resonances remains to be studied. 

The second method is based on  DC-TDHF calculations of the nucleus-nucleus potential.
This potential is shown to vary with the bombarding energy as an effect of the couplings. 
In this approach, all couplings are included to all order, but only in an approximated way.
Indeed, instead of a barrier distribution, the system is sensitive to only one average potential barrier.
Well above the barrier, however, the couplings do not have time to induce a change of the nuclear density and the bare potential is recovered. 

These calculations are the first steps in a series of studies of more and more complicated systems.
Indeed, the couplings to rotational states could be studied in a similar way \cite{sim04,uma08}. 
Applications to asymmetric systems could lead to valuable information on the role of transfer channels. 
Heavier systems will also allow study of the effect of dissipative dynamics on fusion \cite{gol09,ked10,sim11,sim12,guo12,was14}.

It is likely that one will have to go beyond the Hartree-Fock approximation which is used in the present case  to determine the structure properties of the nuclei as well as their potentials and dynamics.
Recent developments including pairing could be used to improve the description of the dynamics of non-magic nuclei \cite{ave08,eba10,ste11,has12,sca12,sca13b}. 
Techniques to compute transfer probabilities have also been developed \cite{sim10,sim11,sek13,was09,sca13b} and could be used to investigate the effect of transfer on fusion.

\end{document}